\documentclass{sig-alternate}
\usepackage{graphicx,url}

\newsavebox{\fmbox}      
\newenvironment{fmpage}[1]
{\begin{lrbox}{\fmbox}\begin{minipage}{#1}}
{\end{minipage}\end{lrbox}\fbox{\usebox{\fmbox}}}  
\newtheorem{definition}{Definition}

\newtheorem{theoreme}{Theorem}
\newtheorem{corollaire}{Corollary}

\newcommand{\shuffle}{\rotatebox{-90}{\hspace{-0.15cm}$\exists$}}

\newcounter{mylistctr}

\begin{document}
\conferenceinfo{RT'06,}{July 20, 2006, Portland, ME, USA}
\CopyrightYear{2006}
\crdata{1-59593-457-X/06/0007}

\title{Uniform Random Sampling of Traces in Very Large Models}

\numberofauthors{5}
%

\author{
%
\alignauthor The RaST group\\
      \affaddr{More informations in section~\ref{auth}}\\
}
\additionalauthors{\label{auth} The RaST group (Random Software Testing) is composed of:
\begin{itemize}
\item Alain Denise - alain.denise@lri.fr\\
LRI, Universit\'e Paris-Sud, UMR CNRS 8623.
\item Marie-Claude Gaudel - mcg@lri.fr\\
LRI, Universit\'e Paris-Sud, UMR CNRS 8623.
\item Sandrine-Dominique Gouraud - gouraud@lri.fr\\
LRI, Universit\'e Paris-Sud, UMR CNRS 8623.
\item Richard Lassaigne - lassaign@logique.jussieu.fr \\
 Equipe de Logique Math\'ematique, Universit\'e Paris VII, UMR CNRS 7056.   
\item Sylvain Peyronnet - syp@lrde.epita.fr\\
LRDE/EPITA and Equipe de Logique Math\'ematique, UMR CNRS 7056, Universit\'e Paris VII.
\end{itemize}}
\date{}
\maketitle
\begin{abstract}
This paper presents some first results on how to perform uniform
random walks (where every trace has the same probability to occur) in
very large models.  The models considered here are described in a
succinct way as a set of communicating reactive modules.  The method
relies upon techniques for counting and drawing uniformly at random
words in regular languages.  Each module is considered as an automaton
defining such a language.  It is shown how it is possible to combine
local uniform drawings of traces, and to obtain some global uniform
random sampling, without construction of the global model.
\end{abstract}



\section{Introduction}


Model based testing has received a lot of attention for years and is
now a well established discipline (see for instance \cite{LY, BT}).
Most approaches have focused on the deterministic derivation from a
finite model of some so-called checking sequence, or of some
complete/exhaustive set of test sequences, that ensure conformance of
the implementation under test ($IUT$) with respect to the model.
However, in very large models, such approaches are not practicable and
some selection strategy must be applied to obtain tests of reasonable
size.  A popular selection criterion is transition coverage.  Other
selection methods rely upon the statement of some test purpose.

With the emergence of model checking, several sophisticated techniques
for the representation and the treatment of models and formulas have
been proposed and used for developing powerful verification tools for
large models.  Among them, one can cite: symbolic model checking,
partial-order reduction methods, reactive modules, symmetry reduction,
hash compaction and bounded model checking.

In this area, several authors have recently suggested the use of
random walks in the state space of very large models in order to get
good approximate checks in cases where exhaustive check is too
expensive \cite{SG03, vmcai, grosu, PHA05}.  This is in the line of
testing methods developed earlier in the area of communication
protocols \cite{West89, MP94, guided, zaidi}.

A random walk \cite{Ald91} in the state space of a model is a sequence
of states $s_0$, $s_1$, $\ldots$ , $s_n$ such that $s_i$ is a state
that is chosen uniformly at random among the successors of the state
$s_{i-1}$, for i = 1, $\ldots$, n.  It is easy to implement and only
requires local knowledge of the model.  In \cite{West89} West reported
experiments where random walk methods had good and stable error
detection power.  In \cite{MP94}, Mihail and Papadimitriou identified
some class of models that can be efficiently tested by random walk
exploration: the random walk converges to the uniform distribution
over the state space in polynomial time with respect to the size of
the model.  These were first evidence of the interest of such
approaches for dealing with special classes of large models.

\begin{figure}
\begin{center}
\includegraphics[width=8cm]{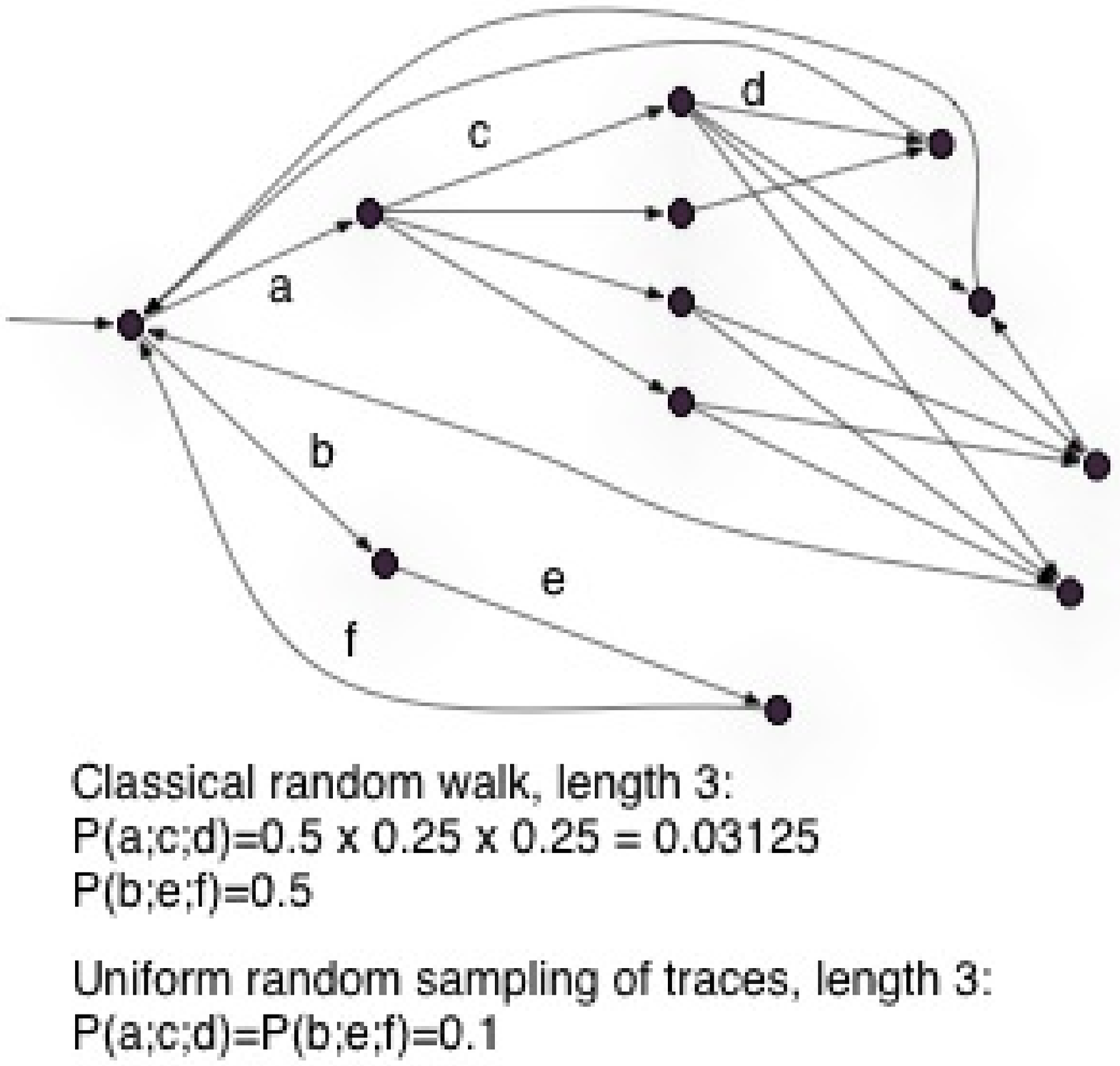}
\end{center}
\caption{The case of irregular topology}
\label{fig:mcg}
\end{figure}

However, as noted by Sivaraj and Gopalakrishnan in \cite{SG03}, random
walk methods have some drawbacks.  In case of irregular topology of
the underlying transition graph, uniform choice of the next state is
far from being optimal from a coverage point of view (see Figure~\ref{fig:mcg}).  Moreover, for
the same reason, it is generally not possible to get any estimation of
the test
coverage obtained after one or several random walks: it would require
some complex global analysis of the topology of the model.  One way to
overcome these problems has been proposed by Gouraud et al.  for
program testing in \cite{GDG01, DGG04}.  It relies upon techniques for
counting and drawing uniformly at random combinatorial structures.
Two major approaches have been developed for dealing with these 
problems: The Markov Chain Monte-Carlo approach (see e.g. the survey 
by Jerrum and Sinclair \cite{JS96}) and the so-called recursive method, as 
described by Flajolet et al. in \cite{Fla} and implemented in \cite{mupad}. 
Although the former is more general in its applications, we chose to work 
with the latter because it is particularly efficient for generating the kind of 
random walks we deal with. 
The idea in \cite{GDG01, DGG04} is to give up
the uniform choice of the next state and to bias this choice according
to the number of elements (traces, or states, or transitions)
reachable via each successor.  Considering the number of traces makes
it possible to ensure a uniform probability on traces.  Considering
elements, such as states or transitions, makes it possible to maximise
the minimum probability to reach such an element.

For addressing very large models, it seems interesting to study how to
combine this improved version of random walk with the representation
techniques developed for struggling against combinatorial state
explosions.  In this paper we present some first results on how to
uniformly sample traces in models described as a set of interacting
transition systems, using the so-called ``reactive modules" notation.
This language, defined by Alur and Henzinger in \cite{fullreactive} is
used as input of the Mocha model checkers and its variants
\cite{Mocha, MochaBis}.

In the probabilistic model checking community, it is the input
language of the PRISM \cite{prism,knp02} and APMC \cite{apmc} model
checkers.  It is similar to communicating extended state machines,
where transitions can be labelled by probabilities.  We propose some
way, inspired from \cite{DGG04}, for uniformly random sampling traces
in systems described by reactive modules, without constructing the
global model.  This method opens interesting perspectives for random
model based testing, for model checking, and for simulation methods.

The paper is organised in two parts. 

In Section 2, we first describe in \ref{react} the reactive modules
notation; then, in \ref{random}, we show how to implement classical
random walk in systems described by reactive modules; in 2.3 we give
an approximation of the detection power of such methods.

In Section 3 we address the computation of probabilities for improving
random walk by uniformly drawing traces in models given as a set of
such modules: 3.1 and 3.2 recall some results on automata
and on counting and drawing uniformly at random words of a given
length, in regular languages; we generalise these techniques to
shuffles of such languages; 3.3 and 3.4 deal with uniform
generation of traces for systems described by reactive modules,
without, and then with, synchronisation.

\section{Random walks in ``reactive modules"}

Our approach is based on a rather classical kind of model in testing,
namely transition systems where transitions are labelled by atomic
actions of a given language $Act$.

\begin{definition}
  An action-labelled transition system\\ ($ALTS$) is a structure
  $\mathcal{M} = (S,T,s_0,Act)$ where $S$ is a set of states, $s_0$
  the initial state, $T \subseteq S \times Act \times S$ a transition
  relation and $Act$ a set of actions.
\end{definition}

In this paper we consider finite $ALTS$. 
Note that, with this definition, $ALTS$ may be non deterministic: the
transition relation may associate several target states to a given
state and a given action.

\subsection{Reactive Modules}
\label{react}
In this paper, we use 
the Reactive Modules language \cite{fullreactive} for describing
$ALTS$.  This language is used in the probabilistic model checking
community for modeling programs and protocols as transition systems.
Two model checkers are using a subset of it as input language: PRISM
\cite{knp02,prism} and APMC \cite{apmc}.

In this language, transition systems are represented by {\em modules}
that can interact together. Each module is composed of local {\em
variables} and guarded commands.  The global state of the system is
given by the local states (i.e. the values of the local variables) of
the modules. More precisely, at any moment the global state of the
system is represented by a vector containing the values of all the
variables of the system. A guarded command is a description of an
atomic transition. It is written as
\begin{center}
\texttt{[sync] guard -> act1 + ... + actk ;}
\end{center}

where \texttt{guard} is a propositional formula over the variables of
the system and where each action (\texttt{act1},...,\texttt{actk})
defines a new assignment of some local variables. 
The choice of the action to be activated is done non
deterministically among those with a valid guard.

Basically, to compute an execution of the whole system, the algorithm
is the following (when there is no synchronization):
\begin{enumerate}
\item Choose non deterministically one of the modules.
\item Check all the guards of the module, keep a list of the valid
guards.
\item If there is no valid guards, no action can be executed, then 
the execution is stopped (to avoid livelock situation).
\item Choose non deterministically among the valid guards, execute non
deterministically one of the corresponding actions. 
\item Modify the local state, thus inducing a modification of the
global state.
\item Go to step 1.
\end{enumerate}

Moreover, one can see that there is a specific field in the guarded
command: \texttt{[sync]}. This field is used to synchronize
modules. By putting a synchronization between guards of different
modules, we force the actions associated to the guards to be done
together (this is a way to describe succinctly a complex behaviour).
Basically, we have to maintain, together with the valid guards, the
corresponding synchronisations.  At the step 2 of the computation, a
guard $g$ synchronised by $s$ in a module $m$ is considered valid if
and only if the guard is true and if there exists, in each module, at
least one guard which is true and synchronised by $s$. If $g$ is
picked at the step 4, then in each module one of the actions corresponding to
one (choosen non deterministically) of the synchronised valid guard is
executed together with the one of actions of $g$.

\begin{figure}
\begin{center}
\includegraphics[width=7.5cm]{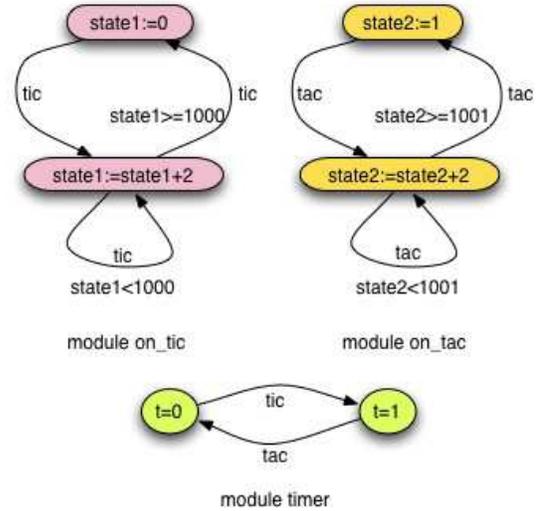}
\caption{Scheme of the example}
\label{fig:RM-fig}
\end{center}
\end{figure}

In the following, we give an example of a simple Reactive
Modules system composed of three modules. All the modules act together
{\em via} synchronization. The figure \ref{fig:RM-fig} summarizes the
example.

\begin{verbatim}

module timer

t : [0..1] init 0;

[tic] t=0 -> t'=1;
[tac] t=1 -> t'=0;
 
endmodule

module on_tic

state1 : [0..1000] init 0;

[tic] state1<1000 -> state1'=(state1+2);
[tic] state1>=1000 -> state1'=0;

endmodule

module on_tac

state2 : [1..1001] init 1;

[tac] state2<1001 -> state2'=(state2+2);
[tac] state2>=1001 -> state2'=1;

endmodule
\end{verbatim}

We now explain quickly the short example. To compute executions of the
model, one has to first pick one of the modules, for instance module
on\_tic. Then the algorithm checks the valid guards. At the beginning,
the variable $state1$ is lower than 1000, so only the first guard is
valid. We have to activate the first guard, but one can see that there
is a synchronization on it: {\tt tic}. So we have to made each module
acting with the two others via a guard synchronised with {\tt tic}. It
means that the only valid execution is to activate the first guard of
the timer and the module on\_tic (there are no guards synchronised
with {\tt tic} in the third module).  So the system starts from the
initial state $(0,0,1)$. It goes from global states of the form
$(0,state1,state2)$ to $(1, state1+2, state2)$, and from global states
of the form $(1,state1,state2)$ to $(0, state1, state2+2)$.
After a while, $state1$ (resp. $state2$) is set to $0$ (resp. $1$) and
the system restart from the initial state $(0,0,1)$.

More informations about Reactive Modules can be found in the paper of
Alur and Henzinger \cite{fullreactive}, that gives a full account of the
semantics, and some correspondence between modules and transition
systems.

The Reactive Modules notation, 
makes it possible to describe huge transitions systems via
synchronised product (\cite{Arnold}).  In practice,
this notation allows to manipulate large systems without being subject
to the exponential blowup of the state space (for instance systems
with more than $10^{30}$ states, see \cite{vmcai}).

Most of the very large models come from the product of several times
the same module. This is the case with classical distributed
algorithms (\cite{pdmc}), real systems/protocols
(\cite{avocs,rivf,JEG99}).

\subsection{Classical random walks }
\label{random}

 An execution path, or a trace in an $ALTS$, is a finite or infinite
 sequence $\sigma = (s_i,a_i,s_{i+1})$ of transitions satisfying: for
 all $i\geq 0$, there exists $a_i \in Act $ such that
 $(s_i,a_i,s_{i+1}) \in T$.


To perform a random walk in a $ALTS$ $\mathcal{M} $, it is
sufficient to have a succinct representation of it, that we call
$diagram_\mathcal{M} $, that allows to generate algorithmically, for
any state $s$, the set of successors of $s$.
An example of such a diagram is a set of reactive modules defining 
a large model $\mathcal{M} $ (as seen above). But OBDD or other representations 
of LTS satisfy this requirement. 
 
The size of such a diagram can be substantially lower than the size of
the corresponding $ALTS$.  Typically, for Reactive Modules, the size
of $diagram_\mathcal{M} $ is poly-logarithmic in the size of
$\mathcal{M}$.

The following function \textbf{Random Walk}\footnote{This classical
algorithm actually defines a so-called ``preset" random walk. For the
distinction between preset and adaptive checking sequences see
\cite{LY}. We give some hints on how to cope with adaptive random
walks in the conclusion. } uses such a succinct representation to
generate a random path of length $k$ and to check if this path leads
to the detection of some conformance error.  We make the simplifying
assumption that there is a reliable verdict that detects an error when
a fault is reached during the execution of the random walk by the
implementation under test (IUT).
 
\begin{center}
\begin{fmpage}{7cm}
\textbf{Random Walk}\\
\textbf{Input: } $diagram_{\cal M},k$\\
\textbf{Output: } samples a path $\pi$ of length $k$ and check conformance on $\pi$
\begin{enumerate}
\item Generate a random path $\pi = (s_0,\dots,s_k)$ such that for
  $i=0,\dots, k-1$, we choose uniformly $s_{i+1}$ among the successors
  of $s_i$.
\item Submit $\pi$ to the IUT. If $\pi$ detects some conformance error then return $1$ else $0$
\end{enumerate}
\end{fmpage}
\end{center}

A drawback of this approach is that we don't know
the probability distribution that it induces on the paths of the model. 
However, it is possible to 
approximate the error detection probability using approximation
techniques for counting problems \cite{kl2}.

\subsection{Randomised approximation scheme}

Many enumeration and counting problems are known to be strongly
intractable. For example, counting the number of elementary paths between two
given nodes in the graph of a transition system is $\sharp
P$-complete. We recall that $\sharp P$ is the complexity class of
functions associated with counting the numbers of solutions of $NP$
decision problems. A classical method to break this complexity barrier
is to approximate counting problems.

We show that we can approximate the error detection probability
 with a simple randomised algorithm. 
A probability problem is defined by giving as input a succinct
representation of a transition system, a property $x$ and as output
the probability measure $\mu (x)$ of the measurable set of execution
paths satisfying this property. We adapt the notion of
randomised approximation scheme \cite{kl2} to probability problems.

\begin{definition}\label{definition:fpras}
  A randomised approximation scheme  for a probability problem
  \cite{vmcai} is a randomised algorithm $\mathcal{A}$ that takes an input $x$ and a
  real number $\varepsilon >0$ and produces a value $A(x,\varepsilon ,\delta)$ such
  that for any $x$, $ \varepsilon >0$, and $\delta >0$:
\begin{center}
$Pr \big( |A(x,\varepsilon ,\delta) -\mu (x)| < \varepsilon  \big) ~ \geq~ 1-\delta $.
\end{center}
 If the running time of $\mathcal{A}$ is polynomial in $|x|$,
 $\frac{1}{\varepsilon}$ and $\log(\frac{1}{\delta})$, $\mathcal{A}$ is
 said to be fully polynomial.
\end{definition}

Let $Paths_k(s_0)$ be the set of execution paths of origin $s_0$ and of depth $k$. 
We generate random paths in the associated probabilistic space
 and compute a random variable $A$ which
 approximates the error detection probability on the paths of depth $k$, $Prob_k [error]$. 
Consider now the random sampling algorithm $\mathcal{GAA}$ designed
for the approximate computation of $Prob_{k} [error]$:

\begin{center}
\begin{fmpage}{7cm}
\textbf{Generic approximation algorithm ${\cal GAA}$}\\
\textbf{Input: } $diagram_{\cal M},k,\varepsilon,\delta$\\
\textbf{Output: } approximation of $Prob_k [error]$\\
$N:= \ln(\frac{2}{\delta})/ 2 \varepsilon ^2$\\
$A:=0$\\
For $i=1$ to $N$ do\\~~~~~~
 $A:=A+ \textbf{Random Walk}(diagram_{\cal M},k)$\\
Return $A/N$
\end{fmpage}
\end{center}

Our approximation will be correct with confidence $(1-\delta)$ after a
 number $N$ of samples polynomial in $\frac{1}{\varepsilon}$ and
 $\log\frac{1}{\delta}$. This result is obtained by using
 Chernoff-Hoeffding bounds \cite{hoeff} on the tail of the
 distribution of a sum of independent random variables.

\begin{theoreme}\label{fpras}(see \cite{wollic}). 
  The generic approximation algorithm ${\cal GAA}$ is a fully
  polynomial randomised approximation scheme 
 for computing $p=Prob_k [error] $ whenever $p \in ]0,1[$.
\end{theoreme}

The property of existence of conformance error detection is monotone:
if it is true for a finite path $\sigma$, then it is also true for
every infinite extension of this path. Let $Prob [error]$ be the error
detection probability in the probabilistic space associated to the set
$Paths(s_0)$ of infinite execution paths of origin $s_0$.  Then the
sequence $(Prob_k [error])_k$ converges to the limit $Prob[error]$.

We can obtain a randomized approximation of $Prob[error]$ by increasing $k$. 
\begin{corollaire}
  The fixed point algorithm defined by iterating the approximation
  algorithm ${\cal GAA}$ is a randomised approximation scheme for the
  probability problem $p=Prob [\psi] $ whenever $p \in ]0,1[$.
\end{corollaire}

The main interest of this randomised approximation scheme is that it
allows some quantification of the error detection power of a random
walk without construction and analysis of the global system.

\section{Improving random walk coverage}

In this section we study how to improve random walk by changing the
random choice of the successors in such a way that traces are
uniformly distributed.  After some preliminaries, we first address the
case of systems described by a set of concurrent, non synchronised
reactive modules, and then we consider the case where there is some
synchronisation. In both cases, we analyse the (intractable)
complexity of explicitly building the product \cite{Arnold} of the
models corresponding to the modules.  Then we propose a much more
efficient alternative, based on the representation of the modules,
hence without explicitely constructing the whole system.

\subsection{From reactive modules to automata}
\label{secAut}

We briefly recall that a finite state automaton $A$ is denoted as a
5-tuple $A=\langle X,Q,q^0,F,\Delta\rangle$ where $X$ is the alphabet,
$Q$ is the finite set of states, $q^0$ is the initial state, $F$ is
the set of final states and $\Delta:Q\times X\rightarrow Q$ is the
state transition relation.  A finite state automaton $A$ defines a
regular language $L$ on the alphabet $X$.

Let $M_1, M_2, \ldots M_r$ be a set of reactive modules, each one
standing for an {\em ALTS}. Each of the $M_i$'s can be represented in a
straightforward way by a finite-state automaton $A_i = \langle
X_i,Q_i,q^0_i,F_i,\Delta_i\rangle$ where
\begin{itemize}
\item each state of $Q_i$ corresponds to a state of $M_i$,
\item any two different transitions are labelled by two different 
letters of $X_i$ (hence the cardinality of $X_i$ equals the numbers of
transitions in $A_i$),\footnote{This is just a way to identify transitions
in order to use their numbers in the following developments. This has no
consequence on the kind of model considered, deterministic or not.}
\item all states are final states (hence $F_i = Q_i$).
\item the $X_i$'s are pairwise disjoint.
\end{itemize}
Consequently, each of the $A_i$'s defines a regular language $L_i$
where each word is in one-to-one correspondence with a trace in the
reactive module.

\subsection{Combinatorial and algorithmic preliminaries}
\label{secComb}

\subsubsection{Automata and word counting}
\label{count}

Let $L$ be a regular language and let $\ell(n)$ be the number of
words of $L$ of length $n$. According to a well known result~(see {\it
e.g.} \cite[Theorem 8.1]{FlSe01}), there exist an integer $N_1$, a
finite set of complex numbers $\omega_1, \omega_2, \ldots,
\omega_k$ and a finite set of polynomials $R_1(n)$, $R_2(n)$, $\ldots$,
$R_k(n)$ such that
\begin{equation} \label{rat}
  n \geq N_1 \ \rightarrow\ \ell(n) = \sum_{j=1}^k R_j(n) \omega_j^n.
\end{equation}
The number $N_1$, as well as the $\omega_j$'s and the $R_j$'s, can be
computed from an automaton of $L$, with an algorithm of polynomial
complexity according to the size of the automaton. Technical details
are given in Appendix 1.

If the automaton of $L$ statisfies certain conditions (see below),
then there is an unique $i$ such that $|\omega_i| > |\omega_j|$ for
any $j \neq i$, and $R_i(n)$ has degree zero, that is $R_i(n)=C$ for
any $n$, where $C$ is a constant. Thus, if we define $\omega = \omega_i$,
the following formula holds, asymptotically:
\begin{equation} \label{asympt}
  \ell(n) \sim C \omega^n.
\end{equation}
This gives a very good estimation of $\ell(n)$ even for rather small
$n$ since, according to Formulas~(\ref{asympt}) and~(\ref{rat}), $C
\omega^n / \ell(n)$ converges to $1$ at an exponential rate.

A simple sufficient condition for Formula~(\ref{asympt}) to hold is:
the automaton is {\em aperiodic} and {\em strongly connected}. An
automaton is aperiodic if, for any sufficiently large $n$, $l(n) \neq
0$. Now, as stated in Section~\ref{secAut}, all the states of any
automaton which represents a reactive module are final states. Thus
any automaton which represents a reactive module is aperiodic.
Concerning strong connectivity, it is satisfied as soon as there is a
reset.  Moreover, it is a sufficient yet not mandatory condition.  For
instance, for satisfying Formula~(\ref{asympt}), in fact it
suffices to have some unique biggest strongly-connected component in
the automaton. Hence, most ``natural'' automata are
such that this formula is satisfied.  Note that in the sequel we use
Formula~(\ref{asympt}) for the automata corresponding to the component
modules.  

\subsubsection{Automata and word shuffling}
\label{secShuffling}

The {\it shuffle} of two words $w, w^\prime \in X^*$, denoted
$w\,\shuffle\,w^\prime$ is the set
$w\,\shuffle\,w^\prime\!=\{w_1w^\prime_1...w_mw^\prime_m |
w_i,w^\prime_i\in X^*, w\!=\!w_1...w_m, w^\prime\!=\!
w^\prime_1...w^\prime_m\}$. For example, $ab\,\shuffle\,cde\,=\,\{
abcde$, $acbde$, $acdbe$, $acdeb$, $cabde$, $cadbe$, $cadeb$, $cdabe$, $cdaeb$,
$cdeab\}$. The shuffle operation is associative and commutative.  It
naturally generalises for languages: the shuffle of two languages
$L_1$ and $L_2$ is the set
\[
L_1\shuffle L_2= \bigcup_{
                   \begin{array}{l}
                       {\scriptstyle w_1\in L_1,}\\ 
                       {\scriptstyle w_2\in L_2} 
                   \end{array}} 
                           w_1\,\shuffle\,w_2
\]
This easily generalises to any finite number $r$ of languages. And the
following property holds: the shuffle of a set of regular languages
is a regular language.
Indeed, let $r > 0$ and let $L_1, L_2, \ldots, L_r$ be $r$ regular languages. 
Let $A_i = \langle X_i,Q_i,q_i^0,F_i,\Delta_i\rangle$ be an automaton
of $L_i$, for any $1 \leq i \leq r$. 
Then the following finite state automaton recognises $L$: 
$A=\langle X, Q, q_0, F, \Delta\rangle$, where
\begin{itemize}
\item $X = X_1 \cup X_2 \cup \ldots \cup X_r$;
\item $Q = Q_1 \times Q_2 \times \ldots \times Q_r$;
\item $q_0 = (q^0_1,q^0_2,\ldots,q^0_r)$;
\item $F = F_1 \times F_2 \times \ldots \times F_r$;
\item $\Delta((q_1,\ldots, q_i,\ldots,q_r),x))=$
\[
\begin{array}{ll}
(\Delta_1(q_1,x),\ldots,q_i,\ldots,q_r) & \mbox{\rm if}\ x \in X_1\\
\ldots\\
(q_1,\ldots,\Delta_i(q_i,x),\ldots,q_r) & \mbox{\rm if}\ x \in X_i\\
\ldots\\
(q_1,\ldots,q_i,\ldots,\Delta_r(q_r,x)) & \mbox{\rm if}\ x \in X_r\\
\end{array}
\]
\end{itemize}
We call this automaton a {\em shuffling automaton} of $L_1, L_2,
\ldots, L_r$.

Now let $\ell_i(k)$ be the number of words of length $k$ belonging to
the language $L_i$. If the $X_i$'s are pairwise disjoint, then the
number of words of length $n$ belonging to $L$ is:
\begin{equation*}
\ell(n) = \!\!\!\!\!\! \sum_{k_1+\cdots +k_r=n} 
             {n \choose k_1, k_2,\ldots, k_r} 
                \ell_1(k_1)\ell_2(k_2)\ldots \ell_r(k_r)
\end{equation*}

Now, suppose that, as in the previous section, all the $L_i$'s are such that
\begin{equation} \label{asymptLi}
  \ell_i(k) \sim C_i \omega_i^{k} 
\end{equation}
where $C_i$ and $\omega_i$ are two constants. Then
\begin{equation}
\label{eqan2}
\begin{array}{lll}
\ell(n) &\sim& 
          C_1 C_2 \ldots C_r \!\!\!\!\!\!\!\! 
             \displaystyle\sum_{k_1+\cdots +k_r=n}  
               \!\!{n \choose k_1,\ldots, k_r}
                \omega_1^{k_1} \ldots \omega_r^{k_r}\\
        &=&
          C_1 C_2 \ldots C_r 
             (\omega_1 + \omega_2 + \ldots + \omega_r)^n\\
\end{array}
\end{equation}

\subsubsection{Uniform random generation of words in a regular
language}\label{words_generation}

First discussed by Hickey and Cohen\cite{HiCo83}, the method for
generating words of regular languages has been improved and widely
generalized by Flajolet and al \cite{Fla}. 
The principle of the generation process is simple: Starting from
state $q_0$, one draws a word step by step; at each step, the process
consists in choosing a successor of the current state and going to
it. 

The problem is to proceed in such a way that only (and all) words of
length $n$ can be generated, and that they are equiprobably
distributed. 
This is done by choosing successors with suitable probabilities. 
Given any state $s$ of the automaton, let $g_m(s)$ denote the number
of words of length $m$ which connect $s$ to any final state $f\in F$. 
Suppose that, at any step of the generation, we are on state $s$ which
has $k$ successors denoted $s_1,s_2,\ldots,s_k$. 
In addition, suppose that $m>0$ transition remain to be done in order
to get a word of length $n$. 
Then the condition for uniformity is that the probability of choosing state
$s_i$ $(1 \leq i \leq k)$ equals $g_{m-1}(s_i) / g_{m}(s)$. 
In other words, the probability to go to any successor of $s$ must be
proportional to the number of words of suitable length from this
successor to any $f$.

So there is a need to compute the numbers $g_i(s)$ for any $0 \leq i
\leq n$ and any state $s$ of the automaton. 
This can be done by using the following recurrence relations:
\begin{equation} \label{eqrec}
\begin{array}{ccll}
        g_0(s) &=& 1 & \mbox{\rm if}\ s\in F \\
               &=& 0 & \mbox{\rm otherwise} \\
        g_i(s) &=& \sum_{s \rightarrow s'}{g_{i-1}(s')} & \mbox{\rm for}\ i>0
\end{array}
\end{equation}
where $s \rightarrow s'$ means that there exists an letter $x\in X$
such as $(s,x,s')\in \Delta$.

Now the generation scheme is as follows:
\begin{itemize}
\item Preprocessing stage: Compute a table of the $g_i(s)$'s for all
$0 \leq i \leq n$ and all states.
\item Generation stage: Draw the word according to the scheme seen above.
\end{itemize}
Note that the preprocessing stage must be done only once, whatever the
number of words to be generated.  Easy computations show that the
memory space requirement is $n \times |Q|$ integer numbers, where
$|Q|$ stands for the number of states in the automaton.  The number of
arithmetic operations needed for the preprocessing stage, as well as
for the generation stage, is linear in $n$.

\subsection{Generating traces of a system of modules without synchronisation.}

Here we focus on the problem of uniformly (that is equiprobably)
generating traces of a given length $n$ in a system of $r$ reactive
modules.  In a first step, we consider that there is no
synchronisation between the $r$ reactive modules $M_i$. 

Each one is
represented by a finite state automaton $A_i = \langle
X_i,Q_i,q^0_i,F_i,\Delta_i\rangle$.
As stated in Section~\ref{secAut}, each of the $A_i$'s defines a
regular language $L_i$ whose words correspond to the traces within
the corresponding module.
Since there is no synchronisation in the system, clearly there is a
one-to-one correspondence between the set of traces of the system and
the words of $L\,=\,L_1\,\shuffle\,L_2 \shuffle\, \ldots
\shuffle\,L_r$. Thus the problem reduces to uniformly generating words
of length $n$ in $L$. We present two different approaches for this
problem and we discuss their complexity issues.

\subsubsection{Brute force method\label{brute1}}

This first approach consists in constructing the {\em shuffling automaton}
that has been defined in Section~\ref{secShuffling}. Then the
classical algorithms for randomly generating words of a regular
language can be processed, as described in Section~\ref{words_generation}.

Let $C_1 = \sum_{0 \leq i \leq r} \mbox{\rm Card}(X_i)$ and $C_2 =
\prod_{0 \leq i \leq r} \mbox{\rm Card}(Q_i)$. The worst-case complexities
of the two main steps of the algorithm are the following.
\begin{enumerate}
\item Constructing the automaton: This step is performed only once, 
whatever the number of traces to be generated. Its worst-case
complexity is $C_1C_2$ in time and space requirements.
\item Generating traces: Using classical algorithms, generating one word 
requires $nC_1$ time requirement, after a preprocessing stage having
worst-case complexity $nC_1C_2$ in time and space. This preprocessing
stage is performed once, whatever the number of traces to be
generated.
\end{enumerate}
Hence the worst case complexity for generating $m$ traces of length
$n$ is $O(nC_1C_2 + mnC_1)$ in time and $O(nC_1C_2)$ in space. This is
linear in $n$, in $m$, in the total size of the alphabets. Since $C_2 =
\prod_{0 \leq i \leq r} \mbox{\rm Card}(Q_i)$, the complexity 
is exponential according to the number of modules. Thus the algorithm
will be efficient only for a small number of modules.

\subsubsection{``On line'' shuffling method\label{shuffling}}

Here we describe an alternative method which avoids constructing the
above automaton. 
We recall that $\ell_i(k)$ is the number of words of length $k$
belonging to the language $L_i$, and $\ell(k)$ is the number of words
of length $k$ belonging to the language $L$.  The method consists
first in choosing at random, with a suitable probability, the length
$n_i$ of each word $w_i$ of $L_i$ which will contribute to the word
$w$ of $L$ to be generated. Then each $w_i$ is generated
independently. Finally, the shuffle operation is processed. We detail
the method just below.

\begin{enumerate}
\item Choose at random a $r$-uple $(n_1,\ldots,n_r)$ with probability
$\Pr(n_1, \ldots, n_r)$ such that
\begin{equation}
\label{eqproba}
\Pr(n_1, \ldots, n_r)\ =\ 
\frac
{
  {n \choose n_1, \ldots, n_r} \ell_1(n_1)\ldots \ell_r(n_r)  
}
{
  \ell(n)
}
\end{equation}

\item For each $0 \leq i \leq r$, draw uniformly a random word $w_i$ 
of length $n_i$ in $L_i$, using the classical algorithm for generating
words of a regular language.

\item Shuffle the $r$ words. This can be done with the following algorithm:
\end{enumerate}
\begin{center}
\begin{fmpage}{\linewidth}
\textbf{Shuffling $r$ words}\\
\textbf{Input: } $r$ words $w_1,\ldots, w_r$, of length $n_1,\ldots, n_r$\\
\textbf{Output: } word $w$ of length $n=\sum_i n_i$ and drawn uniformly among the
set of shuffles of $w_1,\ldots, w_r$.\\
$w \leftarrow \varepsilon$\\
$n \leftarrow \sum_i n_i$\\
while $n>0$ do\\
\hspace*{0.5cm}choose an integer $i$ between $1$ and $r$ with probability $\frac{n_i}{n}$\\
\hspace*{0.5cm}add the first letter of $w_i$ at the end of $w$\\
\hspace*{0.5cm}remove the first letter of $w_i$\\
\hspace*{0.5cm}$n_i \leftarrow n_i-1$\\
\hspace*{0.5cm}$n \leftarrow n-1$
\end{fmpage}
\end{center}

The word $w$ has been generated equiprobably among all the
words of $L$ of length $n$. Regarding complexity issues, clearly the
complexity of step 3 is linear in $n$. The complexity of step 2 is
linear in $n$, in the maximum of ${\rm Card}(X_i)$ and in the maximum
of ${\rm Card}(Q_i)$, in time as well as in space requirements. The
main contribution to the total worst-case time complexity is the
computation of the suitable probabilities by
Formula~(\ref{eqproba}). The space requirement is $O(1)$ but the number
of terms in $\ell(n)$ is exponential in $n$.
However, if the $L_i$'s satisfy the hypothesis of Formula~(\ref{asymptLi}), 
then, by Formula~(\ref{eqan2}): 
\begin{equation}
\label{eqproba2}
\Pr(n_1, \ldots, n_r)\ \sim\ 
\frac
{
  {n \choose n_1, \ldots, n_r} 
       \omega_1^{n_1} \omega_2^{n_2} \ldots \omega_r^{n_r} 
}
{
  (\omega_1 + \omega_2 + \ldots + \omega_r)^n 
}\,.
\end{equation}
There is an easy algorithm for choosing $n_1, \ldots, n_r$ with this
probability without computing it: take the set of integers
$\{1,\ldots,r\}$ and draw a random sequence by picking independently
$n$ numbers in this set in such a way that the probability to choose
$i$ is $\Pr(i) = \frac{\omega_i}{\omega_1 + \omega_2 + \ldots +
\omega_r}$. Then take $n_i$ as the number of occurrences of $i$ in
this sequence.

Well, one could argue that Formula~(\ref{eqproba2}) only provides an
asymptotic approximation of $\Pr(n_1, \ldots, n_r)$ as $n$ tends to
infinity. However, as noticed in Section~\ref{secComb}, the rate of
convergence is exponential, so Formula~(\ref{eqproba2}) is precise
enough even for rather small $n$.
And for really small $n$ (at least when $n<N_1$ in
Formula~(\ref{rat})), $\Pr(n_1, \ldots, n_r)$ can be computed exactly
by Formulas~(\ref{eqrec}) and~(\ref{eqproba}).

In conclusion, for any large enough $n$, the algorithm generates
traces of length $n$ almost uniformly at random. Its overall
complexity is linear according to $n$, to the maximum of ${\rm
Card}(X_i)$ and to the maximum of ${\rm Card}(Q_i)$, in time as well
as in space requirements. 

\subsection{Generating traces in presence of synchronisation.}

Now we suppose that each module contains exactly one synchronised
transition, denoted $\alpha$. Thus, in the global system all modules
must take $\alpha$ at the same time.

Let $A_1, \ldots, A_r$ be $r$ automata, with alphabets $X_1, \ldots,
X_r$, all containing a common synchronisation symbol $\alpha$, such
that $$ \forall i, j \in 1 \ldots r, i \neq j, X_i \cap X_j = \{\alpha\}.$$
Let $S_1, \ldots, S_r$ be the respective languages recognised by
$A_1, \ldots, A_r$. Here, any trace can be represented by a
word belonging to the language $S$ defined as follows:
$S$ is the set of words $w\,\in\,X_1\cup\ldots\cup X_r$ 
such that 
$$
  w = w_0 \alpha w_1 \alpha \ldots w_{m-1} \alpha w_m
$$
where the projection of $w$ onto any $X_i$ belongs to $S_i$.  The
number $m$ is the number of synchronisations during the process: each
of the projections contains exactly $m$ letters $\alpha$ (and,
equivalently, there is no $\alpha$ in any of the $w_i$.)

\subsubsection{Again the brute force approach.}

Here the approach consists in constructing the {\em synchronised}
product of $A_1, A_2, \ldots, A_r$, as follows. Let $X_{i,\alpha} =
X_i \setminus \{\alpha \}$.  The synchronised product~\cite{Arnold} of
$A_1, A_2, \ldots, A_r$ with $\{ \alpha \}$ as synchronisation set is
the finite automaton $A = <X, Q, q_0, F, \delta>$, where
\begin{itemize}
\item $X = X_1 \cup X_2 \cup \ldots \cup X_r$;
\item $Q = Q_1 \times Q_2 \times \ldots \times Q_r$;
\item $q_0 = (q^0_1,q^0_2,\ldots,q^0_r)$;
\item $F = F_1 \times F_2 \times \ldots \times F_r$;
\item $\delta$ is as follows:
$$
\begin{array}{l}
  \Delta((q_1,\ldots, q_i,\ldots,q_r),x)) = \\ 
  \hspace{1cm}                 (\Delta_1(q_1,x),\ldots,q_i,\ldots,q_r)
                                   \ \ \mbox{\rm if}\ x \in X_{1,\alpha},\\
  \hspace{1cm}             \ldots\\
  \hspace{1cm}                 (q_1,\ldots,\Delta_i(q_i,x),\ldots,q_r)
                                   \ \ \mbox{\rm if}\ x \in X_{i,\alpha},\\
  \hspace{1cm}               \ldots\\
  \hspace{1cm}                 (q_1,\ldots,q_i,\ldots,\Delta_r(q_r,x))
                                   \ \ \mbox{\rm if}\ x \in X_{r,\alpha}.\\
  \\
  \Delta((q_1,\ldots, q_i,\ldots,q_r),\alpha)) = \\ 
  \hspace{1cm} \delta_1(q_1,\alpha),\ldots,\delta_i(q_i,\alpha),\ldots,\delta_r(q_r,\alpha))
\end{array}
$$
\end{itemize}
This automaton accepts the language $S$ of synchronised traces.
Once it has been built, the generation process is exactly as in
Section~\ref{brute1}, with the same time and space requirements.


\subsubsection{``On line'' generation of synchronised traces }

Here we sketch an algorithm for almost uniformly generating random
synchronised traces of length $n$, avoiding the construction of the
synchronised product. The approach is similar to the one we described
in Section~\ref{shuffling}, although we must be more careful because
of the synchronisations.
Given that each automaton $A_i$ contains a unique transition labeled
by $\alpha$ (the synchronised transition), let $q_{i,1}$ and
$q_{i,2}$ be the states just before and juste after this transition,
respectively. Now let us define, for each $S_i$, the four following
languages:
\begin{itemize}

\item The {\it beginning language}: $B_i$ is the set of words 
corresponding to the paths which start at the initial state of $A_i$,
which do not cross the $\alpha$ transition, and which stop at $q_{i,1}$.

\item The {\it central language}: $C_i$ is the set of words 
corresponding to the paths which start at $q_{i,2}$, which do not
cross the $\alpha$ transition, and which stop at $q_{i,1}$.

\item The {\it ending language}: $E_i$ is the set of words 
corresponding to the paths which start at $q_{i,2}$, which do not
cross the $\alpha$ transition, and which stop anywhere.

\item The {\it non-synchronised language}: $T_i$ is the set of words 
which start at the initial state of $A_i$, which never cross the
$\alpha$ transition, and which stop anywhere.
\end{itemize}
For any $i$, the language $S_i$ can be defined according to $B_i$,
$C_i$, $E_i$ and $T_i$:
\begin{equation*}
  S_i = B_i . (\alpha . C_i)^* . \alpha . E_i\ \cup\ T_i\,.
\end{equation*}
Thus, if we define $B=\shuffle_{i=1}^r B_i$ (resp. $C=\shuffle_{i=1}^r
C_i$, $E=\shuffle_{i=1}^r E_i$, and $T=\shuffle_{i=1}^r T_i$), we
have:
\begin{equation}
  S = B . (\alpha . C)^* . \alpha . E\ \cup\ T\,.
\end{equation}
Now let $s(n)$ (resp. $s_i(n)$, $b(n)$, $b_i(n)$, $c(n)$, $c_i(n)$,
$e(n)$, $e_i(n)$, $t(n)$, $t_i(n)$) be the number of words of length
$n$ in $S$ (resp. $S_i$, $B$, $B_i$, $C$, $C_i$, $E$, $E_i$, $T$,
$T_i$). Additionally, let $s(n,m)$ be the number of words of $S$ of
length $n$ which contain $\alpha$ exactly $m$ times. Let $w$ be one of
these words. If $m>0$, then $w$ writes
$w=w_0.\alpha.w_1.\alpha.\ldots.\alpha.w_m$ where $w_0 \in B$, $w_i
\in C$ for any $1 \leq i < m$, and $w_m \in E$. Finally, let
$s(n,m,i_0,i_m)$ be the number of such words such that the length of
$w_0$ equals $i_0$ and the length of $w_m$ equals $i_m$. Then we have
\begin{eqnarray} \label{sn}
  s(n) &=& \!\!\sum_{i=0}^{n} s(n,i)\,,
\end{eqnarray}
where
\begin{eqnarray} \label{snm}
  s(n,m) &=& 
       \left\{
         \begin{array}{lr}
         t(n) & \mbox{\rm if $m=0$},\\
         {\displaystyle \sum_{i_0+i_m=0}^{n-m} s(n,m,i_0,i_m) }
                          & \mbox{\rm otherwise},
         \end{array}
       \right.
\end{eqnarray}
and, for $m>0$,
\begin{eqnarray} \label{snmi2}
  s(n,m,i_0,i_m) =
            b(i_0) e(i_m)
         \!\!\!\!\!\!\!\!\!\!\!\!\!\!
                \sum_{\begin{array}{l}
                         {\scriptstyle i_1+\ldots+i_{m-1}=}\\
                         {\scriptstyle n-m-i_0-i_m}
                     \end{array}} 
               \!\!\!\!\!\!\!\!\!\!\!\!\!\!\!
               c(i_1) c(i_2) \ldots c(i_{m-1})\,.
\end{eqnarray}

Now suppose that all the 
the $B_i$'s, the $C_i$'s, the $E_i$'s and the $T_i$'s satisfy
Formula~(\ref{asympt}), that is:
\begin{eqnarray*}
b_i(k) &\sim& C_{b,i} \omega_{b,i}^k\,,\\
c_i(k) &\sim& C_{c,i} \omega_{c,i}^k\,,\\
e_i(k) &\sim& C_{e,i} \omega_{e,i}^k\,,\\
t_i(k) &\sim& C_{t,i} \omega_{t,i}^k\,.
\end{eqnarray*}
Then, similarly to Formula~(\ref{eqan2}), we have:
\begin{eqnarray} 
b(k) &\sim& C_{b,1} \ldots C_{b,r} (\omega_{b,1}+ \ldots +\omega_{b,r})^k\,,\\
c(k) &\sim& C_{c,1} \ldots C_{c,r} (\omega_{c,1}+ \ldots +\omega_{c,r})^k\,,
                   \label{bce}\\
e(k) &\sim& C_{e,1} \ldots C_{e,r} (\omega_{e,1}+ \ldots +\omega_{e,r})^k\,,\\
t(k) &\sim& C_{t,1} \ldots C_{t,r} (\omega_{t,1}+ \ldots +\omega_{t,r})^k
                   \label{bce2}\,.
\end{eqnarray}
Consequently, for $m>0$, 
\begin{equation} \label{snmi}
\begin{array}{l}
s(n,m,i_0,i_m)\ \sim\\
    \ \ \ \ \ \ 
    (C_{b,1} \ldots C_{b,r})
                  (C_{c,1} \ldots C_{c,r})^{m-1}
                  (C_{e,1} \ldots C_{e,r})\\
    \ \ \ \ \ \ \ \ \ \ \ \ 
                (\omega_{b,1}+ \ldots +\omega_{b,r})^{i_0}\\
    \ \ \ \ \ \ \ \ \ \ \ \ \ \ 
                (\omega_{c,1}+ \ldots +\omega_{c,r})^{n-m-i_0-i_m}\\
    \ \ \ \ \ \ \ \ \ \ \ \ \ \ \ \ 
                 (\omega_{e,1}+ \ldots +\omega_{e,r})^{i_m}\,.
\end{array}
\end{equation}
Note that computing $s(n,m,i_0,i_m)$ requires $O(nr)$ arithmetic
operations.

Now we can sketch the algorithm for generating a trace
of length $n$.
\begin{enumerate}

\item Using Formula~(\ref{snmi}), compute $s(n,m,i_0,i_m)$ for 
all $m$ such that $1 \leq m \leq n$ and for all pairs $(i_0,i_m)$ such
that $0 \leq i_0+i_m \leq n-m$. This requires $O(n^3 \times rn) =
O(rn^4)$ arithmetic operations. Then compute $s(n,m)$ for all $m$ such
that $1 \leq m \leq n$, using Formula~(\ref{snm}) and, additionally,
Formula~(\ref{bce2}) when $m=0$. Finally compute $s(n)$ by
Formula~(\ref{sn}). It is worth noticing that this preliminary stage
has to be done only once, whatever the number of traces of length $n$
to be generated. Its overall arithmetic complexity is $O(rn^4)$.

\item Choose $m$, the number of synchronisations, with probability
\begin{equation*}
  \Pr(m) = {\frac {s(n,m)} {s(n)}}.
\end{equation*}
Computing these probabilities requires $O(n)$ arithmetic operations in the worst case. 

\item If $m=0$, then generate uniformly at random a word of length $n$ 
in $T$, with the same algorithm as in Section~\ref{shuffling}.

\item If $m > 0$, then:

\begin{enumerate}

\item Choose the length of $w_0$ and the length of $w_m$ by picking
at random a pair $(i_0,i_m)$ with probability 
\begin{equation*} 
  \Pr(i_0,i_m) = {\frac {s(n,m,i_0,i_m)} 
                        {\sum_{k_0+k_m=0}^{n-m} s(n,m,k_0,k_m)}}\,. 
\end{equation*}
Computing these probabilities requires $O(n^2)$ arithmetic operations in the worst case.

\item Choose the lengths of $w_1, w_2, \ldots, w_{m-1}$ by picking
at random a $(m-1)$-uple $(i_1, i_2, \ldots i_{m-1})$ with probability
\begin{eqnarray*}
  \Pr(i_1, \ldots i_{m-1}) &=& 
        {\frac 
            {c(i_1) c(i_2) \ldots c(i_{m-1})}
            {\sum_P 
                            c(k_1) c(k_2) \ldots c(k_{m-1})}}\,.
\end{eqnarray*}
where $P$ stands for:
\begin{equation*}
  k_1+k_2+\cdots+k_{m-1}=n-m-i_0-i_m. 
\end{equation*}
Using Formula~(\ref{bce}), this reduces to
\begin{eqnarray} \label{Pr2}
  \Pr(i_1, \ldots i_{m-1}) &\sim& {\frac 1 {{n-2-i_0-i_m \choose m-2}}}
\end{eqnarray}
and, similarly to Section~\ref{shuffling}, there is a simple algorithm
for picking $(i_1, i_2, \ldots i_{m-1})$ at random with this
probability. This algorithm is linear according to $n$ and $m$. The
algorithm and the proof of Formula~(\ref{Pr2}) are given in Appendix 2.

\item Now we have got the whole sequence $(i_0, i_1, ..., i_m)$ with a 
suitable probability. It remains to generate the words $w_0 \in B$,
$w_1, w_2, \ldots, w_{m-1} \in C$ and $w_m \in E$, each $w_k$ having
length $i_k$. Each of these words is simply a shuffle of the $r$
languages $(B_i)_{i=1 \ldots r}$ if $k=0$, $(C_i)_{i=1 \ldots r}$ if
$1 \leq k < m$, $(E_i)_{i=1 \ldots r}$ if $k=m$. For each of the
$w_k$'s, the shuffling algorithm given in Section~\ref{shuffling} can
be used.


\end{enumerate}

\end{enumerate}

As remarked above, the first step of the algorithm, in $O(rn^4)$
operations, has to be done only once. After that, the overall
complexity of generating any random trace of length $n$ is quadratic
according to $n$. And, as in Section~\ref{shuffling}, it is linear according
to the maximum of ${\rm Card}(X_i)$ and to the maximum of ${\rm
Card}(Q_i)$, in time as well as in space requirements. Thus we have
defined an efficient way for approximating the uniform coverage in
presence of one synchonisation for any sufficiently large $n$. The
case where there are several synchronisations labelled by different
symbols is more complex but we think it can be addressed with similar
techniques and simplifications. This is the subject of some ongoing
work.

\section{conclusion and perspectives}

One of the main interest of classical random walk is that it can be
performed on large models with a local knowledge only.  However, it
presents some drawbacks, mainly related to the difficulty to estimate,
without analysing the global topology, the test coverage for a given
number of random walk of some given lengths.  In Section 2, we have
shown how it is possible to approximate it via a randomised
approximation scheme.

In the rest of the paper we have described how to perform globally
uniform random walks in very large models described as sets of
concurrent, smaller, models.  By globally uniform random walk, we mean
that the choice of the successor at every step is biased in such a way
that all traces of the global model have equal probability to be
traversed.

A brute force approach is to count the number of paths of the desired
length starting from each successor and to adjust its probability
accordingly. This is feasible via techniques for counting and drawing
uniformly random combinatorial structures.  However, the complexity of
this approach is linear in the number of states of the considered
model.  This makes it feasible for moderately-sized models only.

Then, we have shown how to use local uniform drawings to build
globally uniform random walks, with a complexity that is linear in the
size of the biggest component model.  We use an estimation of the
number of words, but as soon as the length of the random walks is
sufficient, it is a very good approximation as seen in 3.2 (formulas
(1) and (2)).

This method can be used for random testing, model checking, or
simulation of protocols that involve many distributed entities, as it
is often the case in practice.  It ensures a balanced coverage of all
behaviours, even if the topology of the underlying model is irregular.

This work is a first step only.  First, we plan a campaign of
experiments of the method and of some variants of it.  For instance,
instead of uniform coverage of traces, it is possible to consider uniform
coverage of states, or of transitions as it is done in \cite{DGG04}
for testing C programs.

Moreover, results on counting and generating combinatorial structures
are not limited to words of regular languages. They open numerous
perspectives in the area of random testing.  A possibility that is
worth to explore is the test of non deterministic systems via uniform
generation of tree-like behaviours, i.e. some notion of adaptive
random walk inspired from the classical notion of adaptive checking
sequences \cite{LY}.  It would be also interesting to study how the
approach presented here for descriptions by reactive modules could be
transposed to other succinct representations of large models such as
OBDD, symmetry reduction, etc.

\noindent{\bf Acknowledgement.} We thank Radu Grosu for interesting
discussions that have motivated this work.

\section*{Appendix 1: Counting words of rational languages}

Let $L$ be a language on an alphabet $X$, and, for $n \geq 0$, let
$\ell(n)$ be the number of words of $L$ of length $n$. The {\em
generating series} of $L$ is defined as~:
\begin{equation*}
  f(z) = \sum_{n \geq 0} \ell(n) z^n\,.
\end{equation*}
This is a formal power series of one variable $z$ where the
coefficient of $z^n$ equals the number of words of length $n$ in $L$.
According to well-known results (see {\it e.g.} \cite{BeRe87}), if $L$
is a regular language, then its generating series can be expressed as
a rational function
\begin{equation*}
  f(z) = {\frac {N(z)} {D(z)}}
\end{equation*}
where $N$ and $D$ are two polynomials with integer coefficients. This
function is a solution of a system of $m$ linear equations, where $m$ is
the number of states of a deterministic automaton which recognises
$L$.

The number of words of size $n$ mainly depends on the poles of
$f(z)$, that is on the roots of its denominator $D(z)$ (see {\it e.g.}
\cite[Theorem 8.1]{FlSe01}). Precisely, let $\alpha_1, \alpha_2, \ldots,
\alpha_k$ the poles of $f(z)$ and let $\omega_i= 1/\alpha_i$ for any $i$. Then
there exist an integer $N_1$, and $k$ polynomials $R_1(n)$, $R_2(n)$,
$\ldots$, $R_k(n)$ such that
\begin{equation} \label{rat2}
  n \geq N_1 \ \rightarrow\ \ell(n) = \sum_{j=1}^k R_j(n) \omega_j^n.
\end{equation}
where the degree of any $R_j$ equals the multiplicity of its
corresponding pole $\alpha_j$, minus $1$.

As a corollary of the Perron-Frobenius Theorem~\cite[Theorem 8.5 and
Corollary 8.1]{FlSe01}, if the automaton of $L$ statisfies some
conditions (see below), then its generating series has an unique dominant pole,
that is there exists $i$ such that $|\alpha_i| < |\alpha_j|$ for any
$j \neq i$, and this pole has multiplicity $1$.
Hence $R_j(n)$ has degree zero, say $R_j(n)=C$ where $C$
is a constant. Thus we have, asymptotically,
\begin{equation} 
  \ell(n) \sim C \omega_i^n.
\end{equation}
A sufficient condition for the above formula to hold is: the automaton
is strongly connected and aperiodic. However, as noticed in
Section~\ref{count}, there are a number of weaker conditions which
imply it.

\section*{Appendix 2: Proof of Formula~(\ref{Pr2}) and related algorithm}

We have
\begin{eqnarray*}
  \Pr(i_1, \ldots i_{m-1}) &=& 
        {\frac 
            {c(i_1) c(i_2) \ldots c(i_{m-1})}
            {\sum_P 
                            c(k_1) c(k_2) \ldots c(k_{m-1})}}\,
\end{eqnarray*}
where $P$ stands for:
\begin{equation*}
  k_1+k_2+\cdots+k_{m-1}=n-m-i_0-i_m. 
\end{equation*}
By Formula~(\ref{bce}) this leads to
\begin{eqnarray*}
  \Pr(i_1, \ldots i_{m-1}) &\sim&
          {\frac
                {
                    (\omega_{c,1}+ \ldots +\omega_{c,r})^{n-m-i_0-i_m}}
                {\sum_P
                    {
                       (\omega_{c,1}+ \ldots +\omega_{c,r})^{n-m-i_0-i_m}}}}\\
  &=&
  {\frac 1 {\sum_P 1}}.
\end{eqnarray*}
The denominator equals the number of distinct ways to choose
$(k_1,k_2,\ldots,k_{m-1})$ in such a way that they sum to
$n-m-i_0-i_m$. This means that the sequence $(i_1, i_2, \ldots
i_{m-1})$ is to be picked uniformly among all sequences such that
$k_1+k_2+\cdots+k_{m-1}=n-m-i_0-i_m$.

Let $Q=n-m-i_0-i_m$ and $q=m-1$. The number of ways to choose $q$
numbers greater or equal to zero that sum to $Q$ equals $Q+q-1 \choose
q-1$, for any positive integers $Q$ and $q$. Hence
\begin{eqnarray*}
  \Pr(i_1, \ldots i_{m-1}) &\sim&
             {\frac 1 {{n-2-i_0-i_m \choose m-2}}}\,.
\end{eqnarray*}
This proves Formula~(\ref{Pr2}). 

Additionally, there is an easy algorithm to generate uniformly at
random $q$ numbers $i_1$, $i_2$, \ldots, $i_q$ $\geq 0$ that sum to
$Q$: pick uniformly at random $q-1$ numbers $j_1,j_2,\ldots,j_{q-1}$ between
$1$ and $Q+q$, then set $i_1=j_1-1$, $i_2=j_2-j_1-1$, $\ldots$,
$i_{q-1}=j_{q-1}-j_{q-2}-1$, $i_q=Q-j_{q-1}$. Clearly, this simple
algorithm is linear according to $Q$ and $q$, hence to $n$ and $m$.

\clearpage
\balancecolumns 
\end{document}